\documentclass{elsart}
\usepackage{amssymb}
\usepackage{amsmath}
\usepackage{graphicx}
\usepackage{color}

\newcommand{\be}{\begin{equation}}
\newcommand{\ee}{\end{equation}}
\newcommand{\ben}{\begin{eqnarray}}
\newcommand{\een}{\end{eqnarray}}

\begin{document}

\begin{frontmatter}

\title{Finite-size, magnetic and chemical-potential effects on first-order
phase transitions}

\author[UFPA]{E.B.S. Corr\^ea},
\ead{emerson@ufpa.br}
\author[UERJ]{C.A. Linhares},
\ead{linharescesar@gmail.com}
\author[CBPF]{A.P.C. Malbouisson\corauthref{cor2}}
\ead{adolfo@cbpf.br}
%\author[UFBA]{J.M.C. Malbouisson},
%\ead{jmalboui@ufba.br}

\corauth[cor2]{Corresponding author}

\address[UFPA]{ Faculdade de F\'{\i}sica, Universidade Federal do Par\'{a}, 68505-080, Marab\'a, PA, Brazil and Centro Brasileiro de Pesquisas F\'{\i}sicas - CBPF/MCTI,
22290-180, Rio de Janeiro, RJ, Brazil }
\address[UERJ]{Instituto de F\'{\i}sica, Universidade do Estado do Rio de Janeiro, 20559-900, Rio de Janeiro, RJ, Brazil}
\address[CBPF]{Centro Brasileiro de Pesquisas F\'{\i}sicas - CBPF/MCTI,
22290-180, Rio de Janeiro, RJ, Brazil}
%\address[UFBA]{Instituto de F\'{\i}sica, Universidade Federal da
%Bahia, 40210-340, Salvador, BA, Brazil}

\begin{abstract}

We perform a study about effects of an applied magnetic field and a finite
chemical potential on the size-dependent phase structure of a first-order
transition. These effects are introduced by using methods of quantum fields defined on toroidal
spaces, and we study in particular the case
of two compactified dimensions, imaginary time and a spatial one (a heated
film).  It is found that
for any value of the applied field, there is a minimal  size of the
system, independent of the chemical potential, below which the transition disappears. 
\end{abstract}

\end{frontmatter}
{Corresponding author: A.P.C. Malbouisson, tel.: +55-21-21417198; FAX: +55-21-21417556; Email: adolfo@cbpf.br}
\section{Introduction}

On general grounds, systems defined on spaces or spacetimes with some of its
dimensions compactified are of interest in several branches of physics, such
as statistical, condensed-matter, and particle physics. A development of
this kind, which has its roots in the late 1950s, is the systematic approach
to quantum field theory at finite temperature~\cite{3mats1}, as an
imaginary-time formalism. In this formalism, the so-called Matsubara
prescription states that the momentum conjugate to imaginary time is
replaced by frequencies $2n\pi /\beta $ or $2(n+1/2)\pi /\beta $ for bosons
or fermions, respectively, corresponding to the period $\beta =T^{-1}$, with 
$T$ being the temperature. Further developments, as for instance in Refs.~%
\cite{3dj,3ume2,3ume3,3ume4}, allowed to give to the imaginary-time approach
a topological interpretation. It has been shown that the temperature can be
introduced by writing the original theory, formulated in the Euclidean space 
$\mathbf{R} ^{4}$, in the compactified manifold $\Gamma _{4}^{1}=S^{1}\times 
\mathbf{R} ^{3},$ where the compactified dimension is the imaginary time.
The circumference of $S^{1} $ is $\beta $.

An analogous formalism can be constructed for compactified spatial
coordinates, in a $D$-dimensional Euclidean space. In this case, one can
describe systems confined to limited regions of space. This is an idea first
advanced in \cite{birrel1} and we are faced with systems defined on spaces
with compactified spatial coordinates. One takes then a modified Matsubara
prescription in which $\beta $ is replaced by compactification lengths $%
L_{i} $, $i=1,\ldots ,d$, for each bounded spatial coordinate. As is argued
in Ref.~\cite{livro}, this can be interpreted as the system being confined
to a $d$-dimensional parallelepiped embedded in the $D$-dimensional space.
Temperature may be then introduced, as in the Ginzburg--Landau model,
through the mass parameter of the Hamiltonian. By taking $D=3$ and
respectively $d=1$, 2, 3, this can be interpreted as samples of a
superconducting material in the form of a film, a wire, or a grain~\cite%
{livro}.

Since then, progress has been done, in particular to treat jointly spatial
compactification~\cite{birrel1,livro} and the introduction of finite
temperature. Recently, in Refs.~\cite{AOP11} general algebraic foundations
have been presented in this sense, to include concurrently, not only
temperature, but also spatial coordinates, in such a way that any set of
dimensions of the manifold $\mathbf{R}^{D}$ can be compactified. One then
defines a theory in the topology $\Gamma _{D}^{d}=(S ^{1})^{d}\times \mathbf{%
\ R}^{D-d}$, with $1\leq d\leq D$, $d$ being the number of compactified
dimensions. Each of these compactified dimensions has the topology of a
circle and we refer in general to $\Gamma _{D}^{d}$ as a \emph{toroidal}
topology. These ideas, in a simpler form, were already present in Ref.~\cite%
{Ademir} and were applied to the study of spontaneous symmetry
breaking/restoration induced by both temperature and spatial boundaries. In
their more modern presentation, these methods have been recently employed to
investigate several aspects of first and second-order phase transitions in
both bosonic and fermionic systems~\cite%
{Isaque,EPL12(1),luc2,PRD12,AMM2,AMM3}. In this framework, here we intend to
concurrently study effects of a finite chemical potential and of an applied
external magnetic field on the size-dependent phase structure of a
first-order transition. Some physical motivations for such a study are given
along this introductory section, with several references which testify of
the interest of finite-size effects on phase transitions.

In Ref.~\cite{marcio} the Euclidean large-$N$ Ginzburg--Landau model in $D$
dimensions, $d$ of them being compactified, has been considered. The
fixed-point structure of the model is investigated on general grounds, in
the presence of an external magnetic field. An infrared-stable fixed point
has been found, being independent of the number of compactified dimensions,
but for the space dimension $D$ in the range $4<D<6$. This could be related
to studies of extra-dimension effects in both high and low energy physivcs~%
\cite{panilinha,claudio}. In condensed-matter and statistical-physics
contexts, as discussed in Refs.~\cite{moore,moore1} for systems in bulk
form, the fixed point mentioned above should be taken as an indication, not
as a demonstration, of a (formal) continuous transition. This has been
confirmed for a system in the form of a film in Ref.~\cite{malbo}. The
existence of an infrared fixed point in the presence of a magnetic field, as
found in Ref.~\cite{marcio}, does not assure the (formal) existence of a
second-order transition. In any case, for compactified systems under the
action of an external magnetic field, as is also the case for systems in
bulk form, a phase transition for $D\leq 4$, in particular in $D=4$ or $D=3$%
, should not be a second-order one. This furnishes us a motivation to study
first-order phase transitions in the presence of a magnetic background, as
is done in this article. In this sense the present note may be seen as a
extension including concurrently finite-size, magnetic and chemical
potential effects, of previous works on first-order phase transitions that
have already been performed for superconducting materials under the form of
films and wires~\cite{linhares,linhares1}.

There are many potentials that describe first-order transitions both in bulk
and film-like systems, for instance, the Halperin--Lubensky--Ma potential,
engendered by integrating out gauge-field modes~\cite{HLM,claude}. In this
note, we will remain in a somehow less sophisticated framework and will
consider a potential of the Ginzburg--Landau type, $-\lambda \varphi ^4+\eta
\varphi ^6$ ($\lambda >0$, $\eta >0$), which allows that the system
undergoes a first-order transition. However, this study will be done in the
spirit of an application of the above mentioned developments for field
theories defined on toroidal spaces~\cite{AOP11}, including
finite-temperature field theory ideas and compactification of spatial
coordinates, not using the Ginzburg--Landau approximation of considering a
linear behavior of the mass term of the Hamiltonian with the temperature. We
perform a study of concurrent effects of a finite chemical potential and of
an applied external magnetic field, on the size-dependent phase structure of
a first-order transition. Our main concern will be to analyze the model
within a field-theoretical approach, as applied to statistical and
condensed-matter physics. We will consider the particular case of two
compactified dimensions ($d=2$), related to finite temperature and one
compactified spatial coordinate, with compactification length $L$. From a
condensed-matter physical point of view, we can think of this system as a
heated film of thickness $L$, undergoing a first-order phase transition
under the influence of an applied magnetic field.

We remember that Hamiltonian densities, when taken in the Ginzburg--Landau
approximation for temperatures around a given fixed temperature parameter,
are currently employed to describe systems (for instance, superconductors) 
in the absence or the presence of a magnetic background. This has been the
case in which this approximation has been employed to perform studies of
superconducting films in a magnetic field in Refs.~\cite{malbo,malbo1}.
Here, instead of introducing temperature via the mass term, as in the
Ginzburg--Landau approximation, we will consider the system in the framework
of finite-temperature field theory, with $m_0^2$ being a fixed squared mass
parameter; within this formalism, the model is valid for the whole domain of
temperatures, $0\leq T<\infty$.

In this case, we start from the scalar field model described by the
following Hamiltonian density in a Euclidean $D$-dimensional space,
including both $\varphi ^{4}$ and $\varphi ^{6}$ interactions, at zero
temperature, in the absence of boundaries and in the presence of an external
field (in natural units, $\hbar =c=k_{B}=1$): 
\begin{equation}
\mathcal{H}=\left\vert{D}_{\mu }\varphi \right\vert ^2 +m_{0}^{2}\left\vert
\varphi \right\vert ^{2}-\frac{\lambda }{4}\left\vert \varphi \right\vert
^{4}+\frac{\eta }{6}\left\vert \varphi \right\vert ^{6}.  \label{lagrangiana}
\end{equation}%
In the above equation, $m_{0}^{2}$ is a \emph{physical} squared mass
parameter, $\lambda >0$ and $\eta >0$ are, respectively, \emph{physical}
quartic and sextic self-coupling constants, all at zero temperature and in
the absence of spatial compactification; these quantities are taken as fixed
parameters which define the model. Actually, we will define dimensionless
quantities in terms of $m_{0}$, and only $\lambda $ and $\eta $ will be
adjustable parameters. The symbol ${D}$ stands for the covariant derivative, 
${D}_{\mu }=\partial _{\mu }-ieA_{\mu }^{\mathrm{ext}}$, and $A_{\mu }^{%
\mathrm{ext}}$ is an external gauge field.

\section{Zero-temperature magnetic effects in the absence of spatial
boundaries}

In the $D$-dimensional space with Cartesian coordinates $x_1,x_2,\dots x_D$,
following Ref.~\cite{lawrie1}, we choose a gauge such that $A^{ \mathrm{ext}%
}=(0,xH,0,\ldots ,0)$ (to simplify notation we take $x_1\equiv x$), where $H$
is the applied magnetic field, parallel to the $x_3\equiv z$ axis. In this
case, the part of the Hamiltonian $\int d^{D}r\,\mathcal{H}$ quadratic in $%
\varphi $ becomes, after an integration by parts, $-\int d^{D}r\,\varphi
^{\ast} \mathcal{D}\varphi $, where the differential operator $\mathcal{D}$
is 
\begin{equation}
\mathcal{D}=\nabla ^2 -2i\omega x\partial _{y}-\omega ^{2}x^{2}-{m_{0}}^{2},
\label{D}
\end{equation}%
with $\omega =eH$ being the cyclotron frequency. Thus the natural basis to
expand the field operators is the set of the normalized eigenfunctions of
the operator $\mathcal{D}$, the Landau basis, 
\begin{eqnarray}
\chi _{\ell ,p_{y},k}(r) =\frac{1}{\sqrt{2^{\ell }\ell !}}\left( \frac{%
\omega }{\pi }\right) ^{\frac{1}{4}}e^{ik\cdot {\mathcal{Z}}%
}e^{ip_{y}\,y}e^{-\omega (x-p_{y})^{2}/2} H_{\ell }\left[ \sqrt{\omega }%
(x-p_{y})\right] ,  \label{landau}
\end{eqnarray}%
where $r=(x,y,\mathcal{Z})$ and $H_{\ell }$ are the Hermite polynomials; the
corresponding energy eigenvalues are (the subscript $\ell$ denotes the
Landau levels) %\begin{equation}
$E_{\ell }(k)=k^{2}+(2\ell +1)\omega +m_{0}^{2}$; %\label{EigenValue}
%\end{equation}%
$k$ and $\mathcal{Z}$ are conjugate momentum and space $(D-2)$-dimensional
vectors, respectively. The free propagator is written as~\cite{lawrie1} 
\begin{equation}
{\mathcal{G}}(r,r^{\prime })=\int \frac{d^{D-2}k}{(2\pi )^{D-2}}\int
dp_{y}\,\omega \sum_{\ell =0}^{\infty }\frac{\chi _{\ell ,p_{y},k}(r)\chi
_{\ell ,p_{y},k}^{\ast }(r^{\prime })}{k^{2}+(2\ell +1)\omega +m_{0}^{2}}.
\label{Propagator}
\end{equation}
{The non-translational-invariant phase of the propagator (\ref%
{Propagator}) can be isolated as in Ref.~\cite{lawrie1}, and  and we can
write 
\begin{equation}
{\mathcal{G}}({r},{r}^{\prime};\omega) = e^{i \omega (x + x^{\prime}) (y -
y^{\prime}) / 2}\, \bar{\mathcal{G}} ({r} - {r}^{\prime};\omega) ,
\label{GG}
\end{equation}
where $\bar{\mathcal{G}} ({r} - {r}^{\prime};\omega)$ is the translationally
invariant part;  the momentum-space propagator can be obtained from Eqs.~(%
\ref{Propagator}) and (\ref{GG}), by inserting Eq.~(\ref{landau}) into Eq.~(%
\ref{Propagator}) and then considering $r=r^{\prime }$. {This
will be fully justified at the next section, where we will consider
contributions to the effective potential coming from only two kinds of daisy
diagrams, for which we need to consider just the coincidence limit $%
r=r^{\prime }$}}.

Then we write 
\begin{eqnarray}  \label{Propagador k}
{\mathcal{G}}(r,r) =\int \frac{d^{D-2}k}{(2\pi )^{D-2}}\,\int_{-\infty
}^{+\infty }\,dp_{y}\,\left( \frac{\omega }{\pi }\right) ^{\frac{1}{2}%
}e^{-\omega (x-p_{y})^{2}}  \notag \\
\times \sum_{\ell =0}^{\infty }\frac{1}{2^{\ell }\ell !}\left[ H_{\ell }(%
\sqrt{\omega }(x-p_{y})\right] ^{2}\frac{\omega }{k^{2}+(2\ell +1)\omega
+m_{0}^{2}} \equiv \int \frac{d^{D-2}k\,}{(2\pi )^{D-2}}{\mathcal{G}}%
(k,\omega ).  \notag \\
\end{eqnarray}
In the above equation, two of the dimensions are taken into account by the
introduction of the sum over the Landau levels and the incorporation of the
cyclotron frequency; then, by definition, ${\mathcal{G}}(k,\omega )$ is the
free propagator in the remaining ($D-2$)-dimensional momentum space. Using
the orthonormality relations for the Hermite polynomials, %\[
$\int_{-\infty }^{+\infty }\,du\,H_{n}(u)H_{m}(u)\,\exp({-u^{2}})=\sqrt{\pi }
\,2^{n}\,n!\,\delta _{nm}$, %\]%
we obtain straightforwardly the ($D-2$)-dimensional free propagator in
momentum space in the presence of a magnetic field, 
\begin{equation}
{\mathcal{G}}(k,\omega )=\sum_{\ell =0}^{\infty }\frac{\omega }{
k^{2}+(2\ell +1)\omega +m_{0}^{2}};  \label{Propagatork1}
\end{equation}
this is to be used in the ($D-2$)-dimensional space, in an entirely
analogous manner as in dimension $D$ in the absence of a field.

To be more precise, let us remind that, in general, in a $D$-dimensional
non-compactified Euclidean space in the absence of external field, the
Feynman amplitude for a diagram $G$ in a scalar field theory has an
expression of the form (omitting external constant factors and symmetry
coefficients), 
\begin{equation}
A^{(D)}_G(\{p\})= \int \prod _{i=1}^{I}\frac{d^{D}q_{i}}{(2\pi)^D}\prod _{i
=1}^{I}\frac{1}{q_i^2+m_0^2}\prod _{v=1}^{V}\delta \left(\sum_{i}\epsilon
_{vi}q_i\right),  \label{AGgeraldelta}
\end{equation}
where $\{ p \}$ stands for the set of external momenta, $V$ is the number of
vertices, $I$ is the number of internal lines and $q_i$ stands for the
momentum of each internal line $i$. The quantity $\epsilon _{vi}$ is the 
\textit{incidence matrix}, which equals $1$ if the line $i$ arrives at the
vertex $v$, $-1$ if it starts at $v$, and $0$ otherwise. Performing the
integrations over the internal momenta leads to a choice of independent
loop-momenta $\{k_{\alpha}\}$ and we get 
\begin{equation}
A^{(D)}_G(\{p\})= \int \prod _{\alpha =1}^{L}\frac{d^{D}k_{\alpha}}{(2\pi)^D}%
\prod _{i =1}^{I}\frac{1}{q_i^2(\{p\},\{k_{\alpha}\})+m_0^2},
\label{AGgeral}
\end{equation}
where $L$ is the number of independent loops. The momentum $q_i$ is a 
\textit{\ linear} function of the independent internal momenta $k_{\alpha} $
and of the external momenta $\{p\}$. 

This means that, \textit{taking into
account all Landau levels}, calculations of a generic Feynman amplitude for
a \textit{daisy} diagram, can be performed  in the ($D-2$)-dimensional space
using Eq.~(\ref{Propagatork1}),  in an entirely similar way as in the
absence of the external field, \textit{i.e}, performing, from Eq.~(\ref{AGgeral})
in momentum space, for the momentum  integrations over the independent
momenta $k_\alpha\,,\;\alpha=1,2,\ldots, L$ and for the set of propagators
corresponding to the internal lines $i\,,\;i=1,2,\ldots, I$, the
replacements 
\begin{eqnarray}
\int \prod _{\alpha =1}^{L}\frac{d^{D}k_{\alpha}}{(2\pi)^D} \rightarrow \int
\prod _{\alpha =1}^{L}\frac{d^{D-2}k_{\alpha}}{(2\pi)^{D-2}}\,,  \notag
\label{replacement} \\
\prod _{i =1}^{I}\frac{1}{q_i^2\left(\{p\},\{k_{\alpha}\}\right)+m_0^2}
\rightarrow \prod _{i =1}^{I}\sum_{\ell =0}^{\infty }\frac{\omega }{
q_i^2\left(\{p\},\{k_{\alpha}\}\right)+(2\ell +1)\omega +m_{0}^{2}}.  \notag
\\
\end{eqnarray}
This gives, for a generic Feynman amplitude of a daisy diagram, after taking into account the
applied magnetic field, an expression of the form of an integral in the
remaining ($D-2$)-dimensional momentum space, 
\begin{equation}
A^{(D)}_G(\{p\},\omega)= \int \prod _{\alpha =1}^{L}\frac{d^{D-2}k_{\alpha}}{%
(2\pi)^{D-2}}\prod _{i =1}^{I} \left[\sum_{\ell =0}^{\infty }\frac{\omega }{
q_i^2\left(\{p\},\{k_{\alpha}\}\right)+(2\ell +1)\omega +m_{0}^{2}}\right]. 
\label{AGgeral1}
\end{equation}

\section{Effective potential at finite temperature and chemical potential,
in the presence of boundaries, under the action of an external field}

We consider the system under the influence of an external field, at
temperature $\beta ^{-1}$, and we compactify one of the spatial coordinates
(say, $x$) with compactification length $L$. As is argued in Ref.~\cite%
{livro}, this can be considered as a heated system confined to a region of
space delimited by a pair of parallel planes (a film of thickness $L$). As
already noticed, under these conditions the system makes sense for
dimensions $D\geq 4$. Taking into account the prescriptions (\ref%
{replacement}) in a generic dimension, we use $(D-2)$-dimensional Cartesian
coordinates ${\mathcal{Z}}=(\tau ,{x},{\mathcal{W}})$, where $\tau $
corresponds formally to the imaginary-time (inverse-temperature) coordinate, 
$x$ to a spatial coordinate and $\mathcal{W}$ is a ($D-4$) -dimensional
vector. The momentum conjugate to $\mathcal{Z}$ is $k=(k_{\tau },k_{x},{%
\mathcal{Q}})$, ${\mathcal{Q}}$ being a $(D-4)$-dimensional vector in
momentum space. Then we follow the method described in Refs.~\cite{AOP11} in
the particular case $d=2$, to treat jointly finite temperature and
compactification of one spatial coordinate. This amounts to perform a double
Matsubara prescription, one in imaginary time, as is done in
finite-temperature field theory, and an analogous one in the $x$ coordinate.
We also consider a chemical potential $\mu $ associated to the thermal
reservoir. Therefore, the Feynman rules should be modified according to 
\begin{eqnarray}
\int \frac{dk_{\tau }}{2\pi } &\rightarrow &\frac{1}{\beta }\sum_{n_{\tau
}=-\infty }^{\infty },\qquad k_{\tau }\rightarrow \frac{2n_{\tau }\pi }{%
\beta }-i\mu ,  \notag \\
\int \frac{dk_{x}}{2\pi } &\rightarrow &\frac{1}{L}\sum_{n_{x}=-\infty
}^{\infty },\qquad k_{x}\rightarrow \frac{2n_{x}\pi }{L},  \label{Matsubara2}
\end{eqnarray}%
where $L$ is the size of the system and we remind that attention must be
paid to the conditions in Eq.~(\ref{replacement}).

We consider in principle corrections to the mass, 
\begin{equation}
m^{2}(\beta ,\mu ,L,\omega )=m_{0}^{2}+\Sigma (\beta ,\mu ,L,\omega ),
\label{cri1}
\end{equation}%
and coupling constants, $\lambda (\beta ,\mu ,L,\omega )=\lambda _{0}+\Pi
(\beta ,\mu ,L,\omega )$ and $\eta (\beta ,\mu ,L,\omega )=\eta _{0}+\Xi
(\beta ,\mu ,L,\omega )$. Then, a free-energy density of the Ginzburg--Landau
type can be constructed, 
\begin{equation}
{\mathcal{F}}={\mathcal{F}}_{0}+A\,\left\vert \varphi _{0}\right\vert
^{2}+B\,\left\vert \varphi _{0}\right\vert ^{4}+C\,\left\vert \varphi
_{0}\right\vert ^{6},  \label{FreeEnergy}
\end{equation}%
where $A=m^{2}(\beta ,\mu ,L,\omega )$, $B=-\lambda (\beta ,\mu ,L,\omega
)/4 $ and $C=\eta (\beta ,\mu ,L,\omega )/6$ and where $\varphi _{0}$ is the
vacuum expectation value of the field, $\varphi _{0}=\left\langle 0|\varphi
|0\right\rangle $, the classical field. For the sake of simplicity we will
consider only corrections to the mass, the fixed coupling constants $\lambda 
$ and $\eta $ will be taken as the physical ones. 
\begin{figure}[th]
\includegraphics[{height=2.5cm,width=7.5cm}]{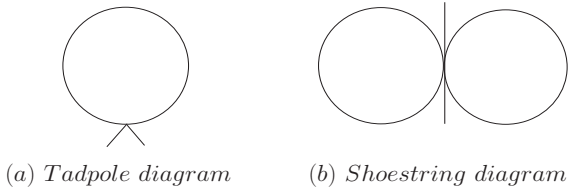}
\caption{Contributions to the effective potential}
\label{figtadshoe}
\end{figure}

Our analysis starts from the effective potential, which is related to the
physical mass through a renormalization condition. In principle, the
effective potential is obtained, following the analysis introduced in Ref.~%
\cite{coleman}, as an expansion in the number of loops in Feynman diagrams.
Accordingly, to the free propagator and to the no-loop (tree) diagrams for
both couplings, radiative corrections are added, with increasing number of
loops. Thus, at the 1-loop approximation, we get the infinite series of
1-loop diagrams with all numbers of insertions of the $\varphi ^4$ vertex
(two external legs in each vertex), plus the infinite series of 1-loop
diagrams with all numbers of insertions of the $\varphi ^6$ vertex (four
external legs in each vertex), plus the infinite series of 1-loop diagrams
with all kinds of mixed numbers of insertions of $\varphi ^4$ and $\varphi
^6 $ vertices. Analogously, we should include all those types of insertions
in diagrams with two loops, etc. This is an extremely hard task; instead of
undertaking this computation, in our approximation we restrict ourselves to
the lowest terms in the loop expansion. The renormalization condition giving
the physical mass then reduces considerably the number of relevant Feynman
diagrams, if we restrict ourselves to first-order terms in both coupling
constants. In this case, just two diagrams need to be considered in this
approximation: a tadpole graph with the $\varphi ^4$ coupling (one loop) and
a ``shoestring'' graph with the $\varphi ^6$ coupling (two loops), as
depicted in Fig.~\ref{figtadshoe}. No diagram with both couplings occur. The
effects of temperature, finite size, chemical potential and magnetic field
appear from the treatment of the loop integrals.

The gap equation we are seeking is given by the condition in which the
physical squared mass is defined as the second derivative of the effective
potential $U(\varphi _{0})$ with respect to the classical field $|\varphi
_{0}|$, taken at zero value, 
\begin{equation}
\left. \frac{\partial ^{2}U(\varphi _{0})}{\partial |\varphi _{0}|{}^{2}}%
\right\vert _{|\varphi _{0}|=0}=m^{2},  \label{renorm}
\end{equation}%
where we remind that $m$ is the \emph{physical} mass. In our case, we will
have a $\beta $, $L$, $\mu $ and $\omega $-dependent squared mass, $%
m^{2}=m^{2}(\beta ,L,\mu ,\omega )$.

Within our approximation, we do not take into account the thermal and
boundary corrections for the interaction coupling constants. As already
stated, they were considered as physical quantities when they were written
in the Hamiltonian at the starting point, being fixed parameters of the
model.\newline

\subsection{The tadpole contribution}

At the one-loop approximation, the contribution from the diagram with only
one $|\varphi _{0}|^{4}$ vertex (the \textit{tadpole}) to the effective
potential, in the presence of a magnetic field, is obtained from the
one-loop contribution to the zero-temperature effective potential in
unbounded space, as an adaptation of the expression in Ref.~\cite{coleman},
taking into account the modified propagator in Eq.~(\ref{Propagatork1}), 
\begin{equation}
U_{1}(\varphi _{0}) =\sum_{s=1}^{\infty }\frac{(-1)^{s+1}}{2s}\left[ \frac{%
\omega \,(-\lambda )|\varphi _{0}|^{2}}{2}\right] ^{s} \int \frac{d^{D-2}k}{%
(2\pi)^{D-2}}\left[\sum_{\ell =0}^{\infty }\frac{1}{\left[ k^{2}+m_{\ell
}^{2}(\omega )\right] ^{s}}\right],  \label{potefet0}
\end{equation}
where we introduce the notation %\begin{equation}
$m_{\ell }^{2}(\omega )\equiv m_{0}^{2}+(2\ell +1)\omega $. 
%\label{massaelle}
%\end{equation}%
As the parameter $s$ counts the number of $\varphi ^{2}$ insertions on the
loop, the tadpole contribution comes from only the $s=1$ term of the sum in
Eq.~(\ref{potefet0}). However, due to analytic continuations that will be
made in the following, the value of $s= 1$ will be taken only at the end of
the calculation.

After changing variables in the integral, $k_{i}/2\pi \rightarrow k_{i}$,
and putting $a_{\tau }=1/\beta ^{2}$, $a_{x}=1/L^{2}$, the one-loop
contribution to the effective potential carrying temperature and finite-size
effects is obtained using Eq.~(\ref{Matsubara2}), as a generalization of
Eq.~(\ref{potefet0}), 
\begin{eqnarray*}  \label{Utad}
U_1(\varphi _0,\beta ,L,\mu,\omega) =\sum_{s=1}^\infty \frac{(-1)^{s+1}}{2s} %
\left[\frac{\omega\,(-\lambda)|\varphi _0|^2}{2}\right] ^s  \notag \\
\times \frac 1{\beta L}\frac 1{\left( 4\pi ^2\right) ^s} \sum_{\ell
=0}^{\infty}\sum_{n_\tau ,n_x=-\infty }^\infty \int \,\frac{d^{D-4}\,{%
\mathcal{Q}}}{\left[ {\mathcal{Q}}^2+a_\tau \left( n_\tau -\frac{i\beta }{%
2\pi }\mu \right) ^2+a_xn_x^2+c_{\ell}^2\right] ^s}\,,  \notag \\
\end{eqnarray*}
where $c_{\ell }^{2}=m_{\ell }^{2}(\omega )/4\pi ^{2}\,.$

The integral in the previous equation is calculated by a
dimensional-regularization formula~\cite{Ramond}, so that the one-loop
contribution to the effective potential can be put into the form

\begin{eqnarray}  \label{potefet5}
U_1(\varphi _0;\beta ,L,\mu,\omega) =\sum_{s=1}^\infty \frac{(-1)^{s+1}}{2s}%
\left[ \frac{\omega\,(-\lambda)|\varphi _0|^2}{2}\right] ^s\frac 1{\beta L}%
\frac{\pi ^{(D-4)/2}}{\left( 4\pi ^2\right) ^s} \frac{\Gamma \left( s-\frac{%
D-4}2\right) }{\Gamma \left( s\right) }  \notag \\
\times \sum_{\ell=0}^{\infty}\sum_{n_\tau ,n_x=-\infty }^\infty \left[
a_\tau \left( n_\tau -\frac{i\beta }{2\pi }\mu \right) ^2+a_xn_x^2+c_{\ell}^2%
\right] ^{(D-4)/2-s}.  \notag \\
\end{eqnarray}
The double sum in Eq.~(\ref{potefet5}) may be recognized as one of the
inhomogeneous Epstein--Hurwitz zeta functions~\cite{Kirsten,Elizalde}, which
gives to the  one-loop contribution to the effective potential the
expression, 
\begin{eqnarray}
U_1(\varphi _0;\beta ,L,\mu ,\omega)&=&\frac {1}{\beta L} \sum_{s=1}^\infty
f(D,s) \left[\frac{\omega\,(-\lambda) |\varphi _0|^2}{2}\right] ^s  \notag \\
&&\times \sum_{\ell=0}^{\infty} Z_2^{c_{\ell}^2}\left( s-\frac{D-4}2;a_\tau
,a_x;b_\tau ,b_x\right) ,  \label{potefet4}
\end{eqnarray}
where $b_\tau =i\beta \mu /2\pi $, $b_x=0$, and 
\begin{equation}
f(D,s)=\frac{\pi ^{(D-4)/2}}{\left( 4\pi ^2\right) ^s}\frac{(-1)^{s+1}}{%
2s\Gamma \left( s\right) }\Gamma \left( s-\frac{D-4}2\right) .
\end{equation}
The zeta functions can be analytically continued to the whole $s$-plane, 
leading to an expression for $Z_{2}^{c^{2}}$ of the general form,

\begin{eqnarray}  \label{Z2}
Z_{2}^{c^{2}}(\nu ;\{a_{j}\};\{b_{j}\}) = \frac{\pi |c|^{2 -
2\nu}\,\Gamma(\nu - 1)}{\Gamma(\nu) \sqrt{a_1 a_2}}  \notag \\
+ \frac{4\pi^{\nu} |c|^{1 - \nu}}{\Gamma(\nu) \sqrt{a_1 a_2}} \left[
\sum_{j=1}^{2} \sum_{n_j = 1}^{\infty} \cos(2\pi n_j b_j) \left( \frac{n_j}{%
\sqrt{a_j}} \right)^{\nu - 1} K_{\nu - 1} \left(\frac{2\pi c n_j}{\sqrt{a_j}}
\right) \right.  \notag \\
\left. + \, 2 \sum_{n_1,n_2=-\infty}^{\infty} \cos(2\pi n_1 b_1) \cos(2\pi
n_2 b_2) \left( \sqrt{\frac{n_1^2}{a_1} +\frac{n_2^2}{a_2}} \right)^{\nu -
1} \right.  \notag \\
\left.\times K_{\nu -1}\left( 2 \pi c \sqrt{\frac{n_1^2}{a_1} + \frac{n_2^2}{%
a_2}} \right) \right] ,  \notag \\
\end{eqnarray}
where $K_{\nu-1}(z)$ are modified Bessel functions of the second kind. For
us, $a_1=a_{\tau}\,,\;a_2=a_x\,, \;b_1=b_{\tau}\,\;, b_2=b_x=0$ and $%
\nu=s-(D-4)/2.$ The first term in Eq.~(\ref{Z2}) is singular for even $D\geq
4$ and will be suppressed by a regularization procedure. This procedure is
known as the zeta-function regularization and is well established, being 
largely employed since a long time in the context of the Casimir effect (see
for instance~\cite{ER});  mathematical foundations for this method are for
instance in~\cite{Elizalde}.

Let us remark that the physical zero-temperature coupling constants in the
absence of boundaries $\lambda$ and $\eta$ have dimensions respectively, of
(mass)$^{4-D}$ and (mass)$^{6-2D}$. We define the dimensionless coupling
constants, $\lambda^{\prime}$, $\eta^{\prime}$; we also define the reduced
temperature $t$, reduced chemical potential $\gamma$, reduced inverse length
of the system $\xi$, and the reduced magnetic field $\delta$, 
\begin{equation}
\lambda^{\prime}=\frac{\lambda}{m_0^{4-D}}\,;\;\;\eta^{\prime}=\frac{\eta}{%
m_0^{6-2D}};\;t =\frac{T}{m_0}\,;\; \xi=\frac{L^{-1}}{m_0}\,\;;\gamma=\frac{%
\mu}{m_0}\;;\delta=\frac{\omega}{m_0^2}\,,  \label{reduzidos}
\end{equation}
in such a way that we have, for any dimension $D$, the set of dimensionless
parameters $\lambda^{\prime}$, $\eta^{\prime}$, $t$, $\gamma$, $\xi$ and $%
\delta$.

In terms of the dimensionless quantities above, after suppression of the
singular term, putting $s$ equal to 1, and using the symmetry property of
Bessel functions $K_{\alpha} (z)=K_{-\alpha }(z)$, the tadpole contribution
to the effective potential is given by 
\begin{equation}
\widetilde{U}_1(\varphi _0;t,\xi,\gamma,\delta ) =-\frac{ \lambda
^{\prime}\delta m_0^2|\varphi _0|^2}{2\left( 2\pi \right) ^{\frac{D-2}{2}}}{%
\mathcal{K}}(t,\xi,\gamma,\delta ),  \label{U1ren}
\end{equation}
where 
\begin{eqnarray}  \label{K1}
{\mathcal{K}}(t,\xi,\gamma,\delta )=\sum_{\ell=0}^{\infty}\left[
\sum_{n=1}^\infty \cosh \left( \frac{\gamma n}{t}\right)\left(\frac {t\sqrt{%
1+(2\ell+1)\delta}}{n}\right) ^{\frac{D-4}{2}} \right.  \notag \\
\left.\times K_{\frac{D-4}{2}}\left( \frac{n\sqrt{1+(2\ell +1)\delta}}{t}%
\right) \right.  \notag \\
\left. +\sum_{n=1}^\infty \left(\frac {\xi \sqrt{1+(2\ell+1)\delta}}{n}%
\right) ^{\frac{D-4}{2}} K_{\frac{D-4}{2}}\left( \frac{n\sqrt{%
1+(2\ell+1)\delta}}{\xi}\right)\right.  \notag \\
\left. +2\sum_{n_1,n_2=1}^\infty \cosh \left( \frac{\gamma n_1}{t}\right)
\left( \frac {\sqrt{1+(2\ell+1)\delta}}{ \sqrt{\frac{n_1^2}{t^2}+\frac{n_2^2%
}{\xi^2}}}\right) ^{\frac{D-4}{2}} \right.  \notag \\
\left.\times K_{\frac{D-4}{2}}\left(\sqrt{\frac{n_1^2}{t^2}+\frac{n_2^2}{%
\xi^2}} \sqrt{1+(2\ell+1)\delta}\right) \right] .  \notag \\
\end{eqnarray}
Notice that in dimension $D=4$, the above expression is well-defined for the
reduced chemical potential restrained to the domain $0\leq \gamma <1$.
Indeed, using an asymptotic formula for large values of the argument $z$ of
the Bessel function, %\begin{eqnarray}
$K_{0}(z)\approx \sqrt{(\pi/2z)}\,\exp(-z)$\,, %\label{bessel0}
%\end{eqnarray}
with $z=(n/t)\sqrt{1+(2\ell +1)\delta}$, we can see that, for large values
of $n$, the argument of the first sum in Eq.~(\ref{K1}) has the asymptotic
form, for arbitrary values of the reduced applied field, 
\begin{eqnarray}  \label{assintotica}
f_{n}(t,\gamma,\delta)\approx \frac{\sqrt{\pi t}}{\sqrt{2\,n\,\sqrt{1+(2\ell
+1)\delta}}} \frac{1}{2}\left[\exp\left(-\frac{n\left(\sqrt{1+(2\ell
+1)\delta}-\gamma\right)}{t}\right)\right.  \notag \\
\left.+\exp\left(-\frac{n\left(\sqrt{1+(2\ell +1)\delta}+\gamma\right)}{t}%
\right) \right].  \notag \\
\end{eqnarray}
The second term inside the square brackets of the above equation does not
present any problem for the convergence of the sum over $n$ for all values
of $\gamma \geq 0$, but the first one implies that the sum over $n$ can be
convergent only if %\begin{equation}
$0\leq \gamma <\sqrt{1+(2\ell +1)\delta}$\,. %\label{gamma}
%\end{equation}
In order to include arbitrarily small values of $\delta$, we should restrain 
$\gamma $ to the domain $0\leq \gamma <1$. A similar argument applies for
the last term in Eq.~(\ref{K1}).

\subsection{The shoestring contribution}

The two-loop shoestring diagram contribution to the effective potential is
obtained using again the Matsubara-modified Feynman rule prescription for
the compactified dimensions. In the absence of boundaries, at zero
temperature, and not submitted to the action of an external field, the
shoestring diagram contribution is simply given by the product, with the
proper coefficients, of two tadpoles, 
\begin{equation}
\widetilde{U}_{2}(\varphi _{0})=\frac{\eta |\varphi _{0}|^{2}}{16}\left[
\sum_{\ell =0}^{\infty }\int \frac{d^{D-2}q}{\left( 2\pi \right) ^{D-2}}%
\frac{1}{\mathbf{q}^{2}+m_{\ell }^{2}(\omega )}\right] ^{2}.
\end{equation}%
Then, after steps analogous to those which have been done for $\widetilde{U}%
_{1}$, we have 
\begin{equation}
\widetilde{U}_{2}(\varphi _{0};t,\xi ,\gamma ,\delta )=\frac{\eta ^{\prime
}\delta ^{2}m_{0}^{2}|\varphi _{0}|^{2}}{4(2\pi )^{D-2}}\left[ {\mathcal{K}}%
(t,\xi ,\gamma ,\delta )\right] ^{2}.  \label{U2ren}
\end{equation}

\subsection{Critical temperature}

We now take $m^{2}(t,\xi ,\gamma ,\delta )\equiv m^{\prime \,2}(t,\xi
,\gamma ,\delta )$ as \textit{dimensionless}, measured in units of $m_{0}^{2}
$. It is obtained from the condition (\ref{renorm}) by using Eq.~(\ref%
{reduzidos}), that is, with the dimensionless coupling constants $\lambda
^{\prime }$, $\eta ^{\prime }$ and in terms of the reduced temperature,
inverse size, chemical potential and external field. At the first order in
the coupling constants $\lambda ^{\prime }$ and $\eta ^{\prime }$ it is
given by 
\begin{equation}
m^{\prime \,2}(t,\xi ,\gamma ,\delta )=\left. \frac{\partial ^{2}}{\partial
\left\vert \varphi _{0}\right\vert ^{2}}\widetilde{U}(\varphi _{0};t,\xi
,\gamma ,\delta )\right\vert _{\left\vert \varphi _{0}\right\vert =0},
\label{cond1}
\end{equation}
where $\widetilde{U}=\widetilde{U}_{0}+\widetilde{U} _{1}+\widetilde{U}_{2}$
and $\widetilde{U}_{0}$ stands for the tree-level approximation. Then, from
Eqs.~(\ref{U1ren}), (\ref{K1}), (\ref{U2ren}), and (\ref{cond1}), we have 
\begin{eqnarray}
m^{\prime \,2}(t,\xi ,\gamma ,\delta )=1-\frac{\lambda ^{\prime }\,\delta }{%
\left( 2\pi \right) ^{\frac{D-2}{2}}}{\mathcal{K}}(t,\xi ,\gamma ,\delta ) +%
\frac{\eta ^{\prime }\,\delta ^{2}}{2(2\pi )^{D-2}}\left[ {\mathcal{K}}%
(t,\xi ,\gamma ,\delta )\right] ^{2}.  \label{massa2}
\end{eqnarray}

As the temperature is lowered, the system approaches the symmetry-breaking
region. Taking the full Eq.~(\ref{massa2}), with $\eta^{\prime} > 0$ and $%
\lambda ^{\prime}> 0$, there is a possibility that the system undergoes a
first-order phase transition. Besides these conditions, it is required that
the minimum values of the free-energy density given by Eq.~(\ref{FreeEnergy}), 
\begin{eqnarray}
{\mathcal{F}} = {\mathcal{F}}_0 + m^{\prime \,2}(t,\xi ,\gamma ,\delta
)\,\left\vert \varphi_0 \right\vert ^{2} -\lambda^{\prime}(\xi ,\gamma
,\delta)\,\left\vert \varphi_0 \right\vert ^{4} + \eta^{\prime}(t,\xi
,\gamma ,\delta)\,\left\vert \varphi_0 \right\vert ^{6},  \label{FreeEnergy1}
\end{eqnarray}
which occur for $\varphi_0$ satisfying $\eta ^{\prime}|\varphi _0|^5 -
\lambda^{\prime} |\varphi _0|^3 + m^{\prime \,2}|\varphi _0| = 0$, should be
equal to $\mathcal{F}_0$, which can be fixed as zero without loss of
generality; this leads to the critical condition, 
\begin{equation}
m^{\prime \,2}(t_{c},\xi ,\gamma ,\delta )=3(\lambda ^{\prime })^{2}/32\eta
^{\prime},  \label{condicao}
\end{equation}
where the mass term is given by the full expression, Eq.~(\ref{massa2}),
containing mass corrections at the first-order in $\lambda^{\prime}$ and $%
\eta^{\prime}$.

The solution of Eq.~(\ref{condicao}) gives the reduced critical temperature $%
t_{c}$ as a function of the reduced inverse size, chemical potential and
applied field,  $t=t_{c}(\xi ,\gamma ,\delta )$.

\section{Magnetic and chemical potential effects on the size-dependent phase
structure: Comments and conclusions}

We fix ourselves in dimension $D=4$. This corresponds to a heated film under
the influence of an external field. In order to perform a qualitative
analysis of the phase structure of the model, we take for the coupling
constants the numerical values $\lambda ^{\prime }=0.5$ and $\eta ^{\prime
}=0.05$. Our objective is to investigate the interplay of the simultaneous
influences of a finite chemical potential and of an applied magnetic field
on the critical temperature as a function of the size of the system. 
%%%%%%%%%%%%%%%
\begin{figure}[htbp]
\begin{minipage}[b]{0.45\linewidth}
    \centering
    \includegraphics[width=\linewidth]{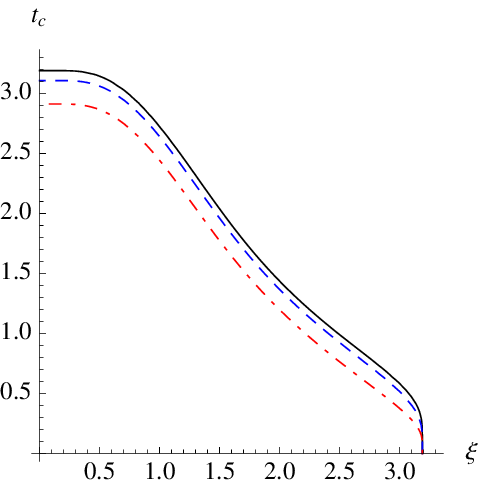}
    \caption{Reduced critical temperature as a function of the reduced inverse
size of the system for dimension $D=4$, for the value of the reduced
magnetic field $\protect\delta =1.5$. We fix $\protect\lambda ^{\prime }=0.5$
and $\protect\eta ^{\prime }=0.05$ and take for the reduced chemical
potential the values $\protect\gamma =0.0$ (full line), $0.5$ (dashed line)
and $0.9$ (dot-dashed line). }
    \label{FigPRE2}
  \end{minipage}
\hspace{0.45cm}  
\begin{minipage}[b]{0.5\linewidth}
    \centering
    \includegraphics[width=\linewidth]{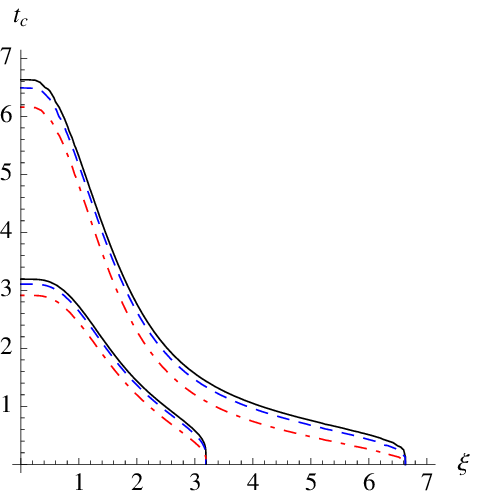}
    \caption{The same as in Fig.~2 for several couples of values $(\protect\gamma ,\protect%
\delta )$. We take for them the values $(0.0,0.3)$, $(0.5,0.3)$ and $%
(0.9,0.3)$ (respectively, full, dashed and dot-dashed lines in the right set
of curves); $(0,1.5)$, $(0.5,1.5)$ and $(0.9,1.5)$ (respectively, full,
dashed and dot-dashed lines in the left set of curves). }
    \label{FigPRE4}
  \end{minipage}
\end{figure}
%%%%%%%%%%%%%%%%

{Let us remind that an effect of the external field is of breaking the translational
symmetry on two of the space dimensions, $x$ and $y$, leaving a ($D-2$)-dimensional translationally invariant subspace. Nevertheless, our system
remains defined on a $D$-dimensional space, although it is not globally
translationally invariant. On the other hand, the general formalism of field
theories in toroidal topologies is constructed for translationally invariant
spaces. This has as a consequence that, if we want to introduce, in the
framework of field theories defined on toroidal spaces~\cite{AOP11}, finite
temperature with chemical potential and finite-size effects, we should
compactify two of the dimensions remaining in the ($D-2$)-dimensional
subspace, in such a way that the theory will be valid for dimensions $D\geq 4
$. In the case of dimension $D=4$, the dimensions of the whole space are $%
\{x,y,z,\tau \}$. We take one of these dimensions, $\tau $, corresponding,
after compactification, to inverse temperature, the three others being
spatial dimensions; then compactification of $\tau $ and of the $z$%
-coordinate (with compactification length $L$) makes our system, embedded in 
\emph{three} spatial dimensions, have the form of a heated film of \textit{%
finite} thickness $L$, under the influence of an applied magnetic field.
Moreover, as we will see below, for each value of the (reduced) applied field, $\delta$,  
the thickness of the film has a lower bound, $L_0(\delta)$, 
sustaining the transition, below which the transition disappears. 

One may speculate on the physics below $L_{0}(\delta)$. Due to the symmetries of the
problem in the two-dimensional limit achieved by taking $L\rightarrow 0$, it
would be expected that the system could have the conditions for a
Berezinsky--Kosterlitz--Thouless (BKT) transition. Actually, numerical calculations show that
the equation for the critical temperature has no solution for $L<L_{0}(\delta)$, and
so we are unable, within our formalism, to investigate this range of
thicknesses. This means, in particular, that we cannot take the $ 
L\rightarrow 0$ limit, in order to verify whether a BKT transition occurs.

In a simpler situation of a first-order transition in the absence of an
applied magnetic field, a similar result was found by some of us in Ref. 
\cite{linhares1} in the context of a Ginzburg--Landau model. In this case,
we have obtained an analytical expression for the critical temperature of  
a superconducting film, as function of its thickness.  
 We found that our predicted curve for the critical temperature is in a relatively good
agreement with experimental data, particularly for small film thicknesses. 
Both the theoretical curve and the experimental data suggest the existence
of a minimal allowed thickness, below which no transition occurs. }

In Fig.~\ref{FigPRE2} we show the reduced critical temperature as a function
of the reduced inverse size of the system, for several values of the reduced
chemical potential at a fixed value of the reduced applied field, $\delta
=1.5 $. We find, in particular, that there exists a minimal size of the
system, $L_{0}$ (corresponding to a maximal reduced inverse size $\xi
_{0}\approx 3.20 $), sustaining the existence of the transition. This
minimal allowed size appears to be \textit{independent} of the value of the
chemical potential.

In Fig.~\ref{FigPRE4} we plot the reduced critical temperature as a function
of the reduced inverse size of the system, for several couples of values $%
(\gamma ,\delta )$. We take for such couples the values $(0,0.5)$, $%
(0.5,0.5) $, and $(0.9,0.5)$; $(0,1.5)$, $(0.5,1.5)$, and $(0.9,1.5)$. We
can infer from this figure that the pattern of Fig.~\ref{FigPRE2} for $%
\delta =0.5$, is reproduced for all values of the reduced applied field. For
each couple of values $(\gamma ,\delta )$ of the (reduced) chemical
potential and applied field, there exists a minimal allowed size of the
system, $L_{0}(\gamma ,\delta )$ [corresponding to a maximal reduced inverse
size $\xi _{0}(\gamma ,\delta )$], below which there is no transition.

From Fig.~\ref{FigPRE4} we can also see that the minimal allowed size of the
system, $L_{0}(\gamma ,\delta )$ [or the maximal allowed value of the
reduced inverse size, $\xi _{0}(\gamma ,\delta )$], is \textit{independent}
of the chemical potential for both values of the reduced applied field, $%
\delta=0.3$ and $\delta=1.5$. Actually, this conclusion is valid for all
values of $\delta$. This is not a trivially expected feature, but we can
prove it by finding the solutions for $\xi _{0}(\gamma ,\delta )$ directly
from Eqs.~(\ref{K1}) and (\ref{condicao}) considering the limit $%
t\rightarrow 0$. Indeed, it should be noted that for $\xi =\xi _{0}(\gamma
,\delta )$, the symmetry-breaking region disappears completely and we have a
null critical temperature. Then $\xi _{0}(\gamma , \delta )$ is obtained by
solving Eq.~(\ref{condicao}) for $t=0$ using an argument similar to one
that was used above to determine the allowed range of values of the reduced
chemical potential. For $t\rightarrow 0$, we use again the asymptotic
formula for large values of the argument of the Bessel function, $%
K_{0}(z)\approx \sqrt{(\pi/2z)}\,\exp(-z)$, for $t\rightarrow 0$, so that
the argument of the first sum between square brackets in Eq.~(\ref{K1}), for
small temperatures is formally the same as in Eq.~(\ref{assintotica}), 
\textit{i.e.}, in this case, the sum can be written as %\[
$\sum_{n=0}^{\infty}f_{n}(t,\gamma,\delta)$. %\]
Taking into account the condition $0\leq \gamma <1$, this sum vanishes in
the limit $t\rightarrow 0$. A similar argument applies for the last term in
Eq.~(\ref{K1}). Therefore, in the limit $t\rightarrow 0$, only the second
term in Eq.~(\ref{K1}) survives, in such a way that all dependence coming
from the chemical potential drops out. Consequently, the resulting solution
of Eq.~(\ref{condicao}), for $\xi$ in this case, $\xi =\xi _{0}(\gamma
,\delta )=\xi_{0}(\delta)$, does not have any influence from the
chemical-potential magnitude.

{The same kind of ``mathematical phenomenon" is found in the
absence of an external field for both first- and second-order phase
transitions~\cite{Isaque,EPL12(1)}. As explicitly stated by some of us in
Ref.~\cite{Isaque}, what appears to happen is that for zero temperature the
 behavior of the physical system  having the minimal (finite) size collapses to the one
corresponding to a zero chemical potential, as is the case of a
Bose--Einstein distribution. In the presence of a magnetic field, for each
value of $\delta$, there is a limiting smallest size of the system, $L_{
\mathrm{min}}(\delta)$, corresponding to a largest reduced inverse size $
\chi_{\mathrm{max}}(\delta)$, over which the first-order transition described by the adopted model, 
ceases to exist. In
other words, in the presence of a magnetic field, we find the same kind of
``mathematical phenomenon", of the collapsing of the system into a
Bose--Einstein distribution for the minimal allowed film thickness; the main difference is 
that the minimal
allowed size for the system is now dependent on the intensity of the applied
field; the larger the field is, the larger is the minimal allowed size of the system.}

Moreover, let us remind that with our choice of gauge for $D=4$, $%
A=(0,xH,0,0)$, the applied field lies on a direction perpendicular to the
film. In this case, we see from Fig.~\ref{FigPRE4} that, for a higher
applied field, the minimal allowed thickness of the film is larger, that is,
thinner films cannot be made for stronger values of the applied field. On
the other hand, let us consider  any film thickness such that the transition
can exist for both values of the applied field ($0<\xi < 3.20$ in the
figure). We see in this case that the critical temperature is lower for
higher applied field, \textit{i.e.}, the applied field goes against the transition.
This behavior for a system in the form of a film is in agreement with the
observed behavior for systems in bulk form, that is, the applied field tends
to destroy the (for instance, superconducting) transition. In other words,
the tendency of the applied field to destroy the phase transition is a
common feature for materials in bulk form and for films, independently of
its thickness. However, the lowering of the critical temperature for a given
thickness and applied field depends on the density of the material, in such
a way that for higher values of the chemical potential, the material
``resists" less to the destruction of the transition by the magnetic
background.

As an overall conclusion, we can say that some of the above results seem 
\textit{a priori} somehow unexpected, such as the independency of the
minimal size of the system (the minimal film thickness) on density and the
fact that for higher applied fields, the minimal allowed thicknesses of the
film are larger; actually, they are a direct consequence of considering
effects coming from the finite size of the system. Other results, such as
the decreasing of the critical temperature as the magnetic field intensity
grows, go along the lines of known features of superconducting materials in
bulk, under the influence of a magnetic background. In any case, the results
found in this note suggest that magnetic and finite-size effects with finite
chemical potential are relevant for bounded systems and significantly
changes the phase structure with respect to the one for the system in bulk
form. In particular, these actors lead to the appearance of a minimal
allowed size of the system, for each value of the applied field, which is
independent of the chemical potential. On the other side, there are other
aspects in agreement with some observations for materials in bulk form.

\section*{Acknowledgments}

A.P.C.M. thanks CNPq and FAPERJ (Brazil), for partial financial support.

\end{document}